\documentstyle[12pt]{article} 

\newcommand{\beq}{\begin{eqnarray}}
\newcommand{\eeq}{\end{eqnarray}}
\newcommand{\rf}[1]{(\ref{#1})}

\begin{document}

\topmargin 0pt
\oddsidemargin 5mm
\headheight 0pt

\topskip 5mm

\thispagestyle{empty}
      
\begin{flushright}
GSUC-CUNY-PHYS-01-2\\
January 2001
\hfill
\end{flushright}

\begin{center}

\hspace{10cm}

\vspace{15pt}
{\large \bf
EVOLUTION OF FIXED-END STRINGS AND THE OFF-SHELL DISK AMPLITUDE  
}

\vspace{20pt}

{\bf Peter Orland}$^{\rm *}$

The Graduate School and University Center, CUNY\\
and Baruch College, CUNY, New York, NY, U.S.A.\\
orland@gursey.baruch.cuny.edu

\vspace{30pt}

{\bf Abstract}
\end{center}

An exact integral expression is found for the amplitude of a Bosonic string with ends separated by
a fixed distance $R$ evolving over a time $T$ between arbitrary  
initial and final configurations. It is impossible 
to make a covariant subtraction of a divergent quantity 
which would render the amplitude non-zero. It is suggested that this 
fact (and not the tachyon) is responsible for the 
lack of a continuum limit of regularized random-surface models 
with target-space dimension greater than one. It appears consistent, however, to 
remove this quantity by hand. The static potential of Alvarez and Arvis
$V(R)$, is recovered from the 
resulting finite amplitude for 
$R>R_{c}=\pi{\sqrt{\frac{(d-2)\alpha^{\prime}}{3}}}$. For $R<R_{c}$,
we find $V(R)=-\infty$, instead of the usual tachyonic result. A
rotation-invariant
expression is proposed for special cases of the off-shell
disk amplitude. {\em None} of the
finite amplitudes discussed are
Nambu or Polyakov functional integrals, except
through an unphysical
analytic continuation. We argue that
the Liouville field does not decouple in off-shell amplitudes, even when
the space-time dimension is twenty-six. 

\vspace{70pt}

{\noindent ${\overline {\rm * Work\; supported \; by\; NSF\; grant\; no.\;0070991,
\;by\; CUNY\; Collaborative \;grant\; no.}}$ }

\noindent 
91915-00-06
and by PSC-CUNY Research Award Program 
grant
no. 69466-00-29.

\vfill
\newpage
\pagestyle{plain}
\setcounter{page}{1}

\section{Introduction}
\setcounter{equation}{0}
\renewcommand{\theequation}{1.\arabic{equation}}

In this paper, we study amplitudes for string world sheets with 
fixed boundaries. Our motivation is to 
better understand off-shell amplitudes for string theory. Knowledge of
these amplitudes yields the effective action of string
theory at lowest nontrivial order. This in turn may lead to a simpler
covariant formalism of string field theory and perhaps yield insight into
QCD as well.

Direct computation of off-shell amplitudes with Polyakov's path integral is 
difficult, since the ghost degrees of freedom are more complicated 
with Dirichlet rather than with Neumann boundary conditions. The boundary
conditions on the ghost fields themselves was discussed long ago
\cite{alvarez1}. Instead of working directly with the Polyakov path
integral as has been done earlier \cite{path int methods}, \cite{jaskolski}
we will study the
energy operator for a string with fixed endpoints
in a covariant gauge, evolving over a finite time
interval. Unfortunately, this places some restrictions on 
the boundary data. On the other hand, the method
does yield a definite expression.

Our first result is that the 
world-sheet-regularized form of
these amplitudes is ill-defined. In general, the expression 
contains a divergent
piece which cannot be removed by a covariant subtraction. We 
believe that this divergence
is related to the inability of lattice and dynamically 
triangulated strings
to have a proper continuum limit for $d>1$ \cite{adj}. The 
form of this
divergent piece is not universal, however. Therefore, it seems that it may be sensibly
set to zero. When this is done, the 
amplitude decays exponentially with area, with the standard
semiclassical corrections
of L\"{u}scher, Symanzik and Weisz \cite{lsw} (who used a loop-wave
equation. See
reference \cite{dop} for a discussion of the Polyakov action). In
fact, the well-known static potential of Alvarez \cite{alvarez} and Arvis \cite{arvis}
is easily
recovered for $R$ greater than the so called critical radius. Instead
of becoming imaginary below this radius, the potential becomes
negative infinite. This does not mean that the usual
conclusion is wrong, for our method of continuing below
the critical radius is different mathematically from that in
references \cite{alvarez}, \cite{arvis}. This phenomenon
can be understood as
as a
divergence of the string amplitude. 

The expression obtained for the amplitude
is not rotation invariant. If the world-sheet boundaries
lie on a rectangle of size $R\times T$, this expression changes upon interchange of 
$R$ and $T$. This is not surprising; in the appendix, we show
that a similar expression constructed for relativistic particles is not Lorentz
invariant. We suggest the corrected expression for the disk amplitude, by
analogy with the particle case. 

The Liouville field does not decouple in off-shell 
amplitudes even in the critical dimension. The correct 
interpretation of this fact is most likely that if the string degrees of freedom $X^{\mu}$
obey Dirichlet boundary conditions, the central
charge of the world-sheet Fadeev-Popov ghosts is $-2$ instead of $-26$.

The particular geometries studied here are fairly special. They 
are such that Cartesian coordinates $X^{\mu}$ have the property that
the boundary values of $X^{0}$ and $X^{1}$ describe a rectangle. The remaining 
coordinates, $X^{\perp}=X^{2},\dots,X^{d-1}$ are left arbitrary on the boundary. For
such boundaries, an exact expression for the amplitude is found. Setting
$X^{\perp}=0$ and letting one dimension of the rectangle go to infinity may be used to
obtain the Bosonic-string static potential \cite{alvarez,arvis}. We
discuss in Section 7 how the amplitude for slightly more 
general boundary geometries
may be determined.

\section{Off-shell amplitudes}
\setcounter{equation}{0}
\renewcommand{\theequation}{2.\arabic{equation}}

The action of the Bosonic string is
\beq
S=\frac{1}{4\pi \alpha^{\prime}}\int d^{2}\xi \;{\sqrt{g}}g^{ab} \;\partial_{a}X^{\mu}
\partial_{b}X^{\mu}\;, \label{action}
\eeq
where $\xi^{a}$ are the coordinates of the
world-sheet, $a,b=0,1$, $\partial_{a}=\partial/\partial \xi^{a}$
and $\mu,\nu=1,\dots d$.  The functional integral of the disk with fixed
boundary data $X(\varphi)$ with boundary coordinate
$\varphi \in [0,2\pi)$
\beq
\Phi[X]=\int_{X(\varphi)}\!\! {\cal D}X \;e^{-S}
\;.
\label{quantum action principle}
\eeq
If this expression exists, it has the
expansion:
\beq
\Phi[X]\approx \frac{1}{\sqrt {-{\rm det}\delta^{2} S}}e^{-S_{0}}
\;,
\nonumber
\eeq
where $S_{0}$ is the classical action of a solution to the equations of motion
and the prefactor is the reciprocal of the square root of the determinant of the
fluctuation operator. We shall argue that \rf{quantum action principle} is not
well-defined as it stands, but that it is possible to formally define a
string Green's function $\Phi[X]$ anyway. Nonetheless, it is interesting
to determine $S_{0}$ from the boundary data, as was 
done in reference \cite{poly}. Though the on-shell tree amplitudes
for string theory obtained from $S_{0}$ are correct \cite{poly}, we will show
that the off-shell amplitudes are not.

\section{Determination of the classical action}
\setcounter{equation}{0}
\renewcommand{\theequation}{3.\arabic{equation}}

It is possible to obtain $S_{0}$ for Dirichlet data $X^{\mu}(\varphi)$ on
the boundary by solving the equations of motion \cite{poly}. We review this
procedure here (which is not generally known) because 
the result will be important in our discussion
of the complete solution for the off-shell disk amplitude. 

Let us suppose
we know the solution in the interior of the disk 
$X^{\mu}(r,\varphi)$, $g_{ab}(r, \varphi)$. These quantities satisfy the equations of
motion
\beq
\frac{1}{\sqrt g}\partial_{a}{\sqrt g}g^{ab}\partial_{b}X^{\mu}=0\;, \label{laplace}
\eeq
and
\beq
\partial_{a}X^{\mu}\partial_{b}X^{\mu}-\frac{1}{2} g_{ab}g^{cd}
\partial_{c}X^{\mu}\partial_{d}X^{\mu}=0\;. \label{energy-momentum}
\eeq
Once any solution is obtained, by making a conformal transformation on the
metric, it is possible to chose
$r\in [0,a)$, $g_{rr}=e^{\phi}$, $g_{\varphi^{\prime} \varphi^{\prime}}=r^{2}e^{\phi}$
and $g_{r \varphi^{\prime}}=0$. The prime in $\varphi^{\prime}$ is present because
$\varphi$ is mapped to $\varphi^{\prime}$ by the conformal transformation. Then the equation of motion is
Laplace's equation \rf{laplace} which has the solution:
\beq
X^{\mu}=B^{\mu}(re^{i\varphi^{\prime}})+{\bar B}^{\mu}(re^{i\varphi^{\prime}}) \;, \nonumber
\eeq
where $B^{\mu}(z)$ is an analytic function $$B^{\mu}(z)=\sum_{n=0}^{\infty}b^{\mu}_{n}z^{n}$$ satisfying
$B^{\mu}(ae^{i\varphi^{\prime}})+{\bar B}^{\mu}(ae^{i\varphi^{\prime}}) =X^{\mu}(\varphi^{\prime})$. The values
of $b^{\mu}_{n}$ are determined by the boundary
data $X^{\mu}(\varphi^{\prime})=X^{\mu}(a,\varphi^{\prime})$:
\beq
b^{\mu}_{n}=\frac{1}{2\pi a^{n}} \int_{0}^{2\pi} \!\!d\varphi^{\prime} e^{-{\rm i}n \varphi^{\prime}}X^{\mu}(\varphi^{\prime})
\;. \label{coefficients}
\eeq
In terms
of the coefficients $b^{\mu}_{n}$, the action \rf{action}
is, after integration by parts and using the
equation of motion, a pure surface term:
\beq
S=S_{0}=\frac{1}{4\pi \alpha^{\prime}}\int_{0}^{a} \!dr \int_{0}^{2\pi}\!\!d\varphi^{\prime} 
\left[r(\partial_{r}X)^{2}+\frac{1}{r}(\partial_{\varphi^{\prime}}X)^{2}
\right] \nonumber
\eeq
\beq
=\frac{1}{4\pi \alpha^{\prime}}\int_{0}^{a} \!dr \int_{0}^{2\pi}\!\!d\varphi^{\prime} \left[ \partial_{r}(rX^{\mu}\partial_{r}X^{\mu})
+\frac{1}{r}\partial_{\varphi^{\prime}}(X^{\mu}\partial_{\varphi^{\prime}} X^{\mu})\right] \;, \label{surface}
\eeq
where $a$ is the disk diameter (the actual value of which is irrelevant). Only the first term in \rf{surface}
is not zero. The general solution for $X$ into \rf{surface} yields
\beq
S_{0}=\frac{1}{2\alpha^{\prime}} \sum_{n=1}^{\infty} n {\bar b}^{\mu}_{n} b^{\mu}_{n} \;,
\nonumber
\eeq
and substituting \rf{coefficients} gives
\beq
S_{0}=-\frac{1}{8\pi \alpha^{\prime}} 
\int_{0}^{2\pi}\!\!d\varphi^{\prime}_{1}\int_{0}^{2\pi}\!\!d\varphi^{\prime}_{2}
\sum_{n=1}^{\infty} ne^{{\rm i}n\;(\varphi^{\prime}_{1}-\varphi^{\prime}_{2})}
\left\{[X(\varphi^{\prime}_{1})-X(\varphi^{\prime}_{2})]^{2} \right.
\nonumber
\eeq
\beq
\left. -X(\varphi^{\prime}_{1})^{2}-X(\varphi^{\prime}_{2})^{2}
\right\}\;. \nonumber
\eeq
The last two terms are zero. Summation over $n$ is straightforward, giving the result:
\beq
e^{-S_{0}}=\exp -\frac{1}{16\pi \alpha^{\prime}}
\int_{0}^{2\pi} \!\! d\varphi^{\prime}_{1} \int_{0}^{2\pi} \!\! d\varphi^{\prime}_{2}\;\;
\frac{[X(\varphi^{\prime}_{1})-X(\varphi^{\prime}_{2})]^{2}}{\sin^{2}\!
\frac{1}{2}(\varphi^{\prime}_{1}-\varphi^{\prime}_{2})}\;.
\nonumber\eeq
This expression is invariant under diffeomorphisms of
$\varphi^{\prime}$, mapping the interval $[0,2\pi)$ to itself. We are therefore free to map $\varphi^{\prime}$
back to
$\varphi$, obtaining
\beq
e^{-S_{0}}=\exp -\frac{1}{16\pi \alpha^{\prime}}
\int_{0}^{2\pi} \!\! d\varphi_{1} \int_{0}^{2\pi} \!\! d\varphi_{2}\;\;
\frac{[X(\varphi_{1})-X(\varphi_{2})]^{2}}{\sin^{2}\!\frac{1}{2}(\varphi_{1}-\varphi_{2})}\;.
\nonumber
\eeq
\beq
=\exp -\frac{1}{4\pi \alpha^{\prime}}
\int_{0}^{2\pi} \!\! d\varphi_{1} \int_{0}^{2\pi} \!\! d\varphi_{2}
\sum_{j=-\infty}^{\infty} \frac{[X(\varphi_{1})-X(\varphi_{2})]^{2}}{(\varphi_{1}-\varphi_{2}
-2\pi j)^{2}} \label{classical action}
\eeq
Since $S_{0}$ is the value of an extremum of the Polyakov
action, it is the value of an extremum of the Nambu action as
well. Therefore $S_{0}$ is the minimal area of a surface whose
boundary is $X^{\mu}(\varphi)$.

The expression
\rf{classical action} is invariant under diffeomorphisms of the boundary, $\varphi
\longrightarrow F(\varphi)$. In the form of this result presented
by Polyakov, the range of $\varphi$ is extended to the entire real
axis, so that diffeomorphisms become SL(2,${\rm I}\!{\rm R}$) 
transformations \cite{poly}. One finds
that the off-shell expression \rf{classical action} becomes
$$e^{-S_{0}}=\exp -\frac{1}{4\pi \alpha^{\prime}}
\int_{-\infty}^{\infty}  d\varphi_{1} \int_{-\infty}^{\infty}  d\varphi_{2}
\;\frac{[X(\varphi_{1})-X(\varphi_{2})]^{2}}{(\varphi_{1}-\varphi_{2}
)^{2}}$$ 
This expression yields the open-string
Koba-Nielsen amplitudes when extended to the mass shell 
\cite{lecture}. Naively, one expects the Liouville field to decouple 
when $d=26$ and
that \rf{classical action} should be the final expression for $Z[X]$. We will see
that this is not the case.

\section{The string energy operator and subtractions}
\setcounter{equation}{0}
\renewcommand{\theequation}{4.\arabic{equation}}

In this section, the string energy operator, which governs time 
development in target space, is critically reexamined.
Certain conventions are different from Arvis' \cite{arvis}. We 
will quantize in the Schr\"{o}dinger representation. 

We change the notation for the coordinates, specifically 
$\xi^{0}=\tau$ and $\xi^{1}=\sigma$, and write the string degrees
of freedom as $X^{\mu}(\sigma, \tau)$. The
coordinate $\sigma$ lies in the interval $[0,\pi]$, while we treat the coordinate
$\tau$ as the world-sheet
time on world sheets of Minkowski signature. Target space will also be assumed to
have Minkowski signature (in the next section, we Wick-rotate to
Euclidean signature). Unlike
the case of the on-shell string, the world-sheet time $\tau$ is {\it not} identified
with target-space time. 

The boundary conditions taken in most of this paper
are $X^{\perp}(0,\tau)=0$, $X^{\perp}(\pi,\tau)=0$,
$X^{1}(0,\tau)=0$, $X^{1}(\pi,\tau)=R$,
$\partial_{\sigma}X^{0}(0,\tau)=0$
and $\partial_{\sigma}X^{0}(\pi,\tau)=0$. We will discuss slightly more general
boundary conditions in Section 7. 

In these coordinates the action \rf{action} is that of a free field theory 
\beq
S=-\frac{1}{4\pi \alpha^{\prime}}\int d\tau \int_{0}^{\pi} \!\!d\sigma \;
\eta_{\mu \nu}[ (\partial_{\tau}X^{\mu})^{2}-(\partial_{\sigma}X^{\mu})^{2}]\;, \nonumber
\eeq
where $\eta$ has signature $(+,-,\dots,-)$. This action
leads to the momentum density, conjugate to $X^{\mu}$
\beq
{\rm P}_{\mu}(\sigma, \tau)=-\eta_{\mu \nu}  \partial_{\tau} X^{\nu}(\sigma, \tau)\;. \nonumber
\eeq
The energy of the string is not the canonical Hamiltonian, but 
rather the integral of ${\rm P}_{0}$:
\beq
E=-\int_{0}^{\pi} \!\!d\sigma \; {\rm P}_{0}(\sigma, \tau) =\int_{0}^{\pi} \!\!d\sigma\;
\partial_{\tau} X^{0}(\sigma, \tau)\;. \label{energy}
\eeq

The boundary conditions are consistent with the normal-mode expansions:
\beq
X^{\perp}(\sigma, \tau)=2\sqrt{\alpha^{\prime}}\sum_{n=1}^{\infty} {\sqrt{n}}
\sin n\sigma \; X^{\perp}_{n}(\tau)\;, \nonumber
\eeq
\beq
X^{1}(\sigma, \tau)=\frac{R}{\pi} \sigma+2\sqrt{\alpha^{\prime}}\sum_{n=1}^{\infty} {\sqrt{n}}
\sin n\sigma \; X^{1}_{n}(\tau)\;, \nonumber
\eeq
\beq
X^{0}(\sigma, \tau)=X^{0}_{0}(\tau)+2\sqrt{\alpha^{\prime}}\sum_{n=1}^{\infty} {\sqrt{n}}
\cos n\sigma \; X^{0}_{n}(\tau)\;. \nonumber
\eeq

The action \rf{action} in conformal coordinates $\sigma, \tau$, after
Wick rotating to Minkowski signatures and written in term of normal
modes is
\beq
S=-\frac{1}{4\pi \alpha^{\prime}}
\int d\tau \{ \pi(\partial_{\tau}X^{0}_{0})^{2}
+\sum_{n=1}^{\infty} [(\partial_{\tau}X^{0}_{n})^{2}-(\partial_{\tau}X^{1}_{n})^{2}
-(\partial_{\tau}X^{\perp}_{n})^{2}   \nonumber
\eeq
\beq
-n^{2}(X^{0}_{n})^{2}+n^{2}(X^{1}_{n})^{2}+n^{2}(X^{\perp}_{n})^{2}
]\}\;. \nonumber
\eeq
The canonical momenta of $X^{0}_{0}$, $X^{0}_{n}$, $X^{1}_{n}$ and $X^{\perp}_{n}$ ($n>0$)
are
\beq
P^{0}_{0}=-\frac{1}{2\alpha^{\prime}}\partial_{\tau}X^{0}_{0}\;,\;\;
P^{0}_{n}=-\frac{1}{2\pi\alpha^{\prime}}\partial_{\tau}X^{0}_{n}\;,\;\;
P^{1}_{n}=\frac{1}{2\pi\alpha^{\prime}}\partial_{\tau}X^{1}_{n}\;,\;\;
P^{\perp}_{n}=\frac{1}{2\pi\alpha^{\prime}}\partial_{\tau}X^{\perp}_{n}\;, \nonumber
\eeq
respectively. It is clear that $P^{0}_{0}$ is equal to the total
energy $E$, defined in \rf{energy}. The canonical Hamiltonian is
\beq
H=-\alpha^{\prime}E^{2}
-\pi \alpha^{\prime}\sum_{n=1}^{\infty}[(P^{0}_{n})^{2}-(P^{1}_{n})^{2}-
(P^{\perp}_{n})^{2}] \nonumber
\eeq
\beq
-\frac{1}{4\pi \alpha^{\prime}}
\sum_{n=1}^{\infty}n^{2}[(X^{0}_{n})^{2}-(X^{1}_{n})^{2}-
(X^{\perp}_{n})^{2}]\;. \label{Hamiltonian}
\eeq
It is not hard to show that the Poisson bracket of $E(\tau)$ with
the Hamiltonian \rf{Hamiltonian} vanishes. Therefore $E$ is independent of $\tau$.

The energy-momentum conditions \rf{energy-momentum}, after 
rotating to Minkowski signature are
\beq
(\partial_{\tau}X^{0}\pm\partial_{\sigma}X^{0})^{2}-
(\partial_{\tau}X^{1}\pm\partial_{\sigma}X^{1})^{2}-
(\partial_{\tau}X^{\perp}\pm\partial_{\sigma}X^{\perp})^{2}=0 \;. \nonumber
\eeq
In terms of normal modes and their conjugate momenta, these conditions are
\beq
L_{n}\equiv \sum_{j+k=n}\eta^{\mu \nu} \alpha^{\mu}_{j}\alpha^{\nu}_{k}=0 \label{virasoro}
\eeq
where $j,k,n$ are (positive, zero or negative) integers and
\beq
\alpha^{0}_{0}={\sqrt{2}}\alpha^{\prime}E\;,\;\;
\alpha^{1}_{0}=\frac{R}{{\sqrt{2}}\pi}\;,\;\; \alpha^{\perp}_{0}=0\;,\nonumber
\eeq
\beq
\alpha^{0}_{n}={\sqrt{\pi}}\alpha^{\prime}P^{0}_{n}-\frac{{\rm i} n}{2{\sqrt{\pi}}}X^{0}_{n}\;,\;\;
\alpha^{0}_{-n}={\sqrt{\pi}}\alpha^{\prime}P^{0}_{n}+\frac{{\rm i} n}{2{\sqrt{\pi}}}X^{0}_{n}\;,
\;\; (n \ge 1) \nonumber
\eeq
\beq
\alpha^{1}_{n}=-{\rm i}{\sqrt{\pi}}\alpha^{\prime}P^{1}_{n}
+\frac{n}{2{\sqrt{\pi}}}X^{1}_{n}\;,\;\;
\alpha^{1}_{-n}={\rm i}{\sqrt{\pi}}\alpha^{\prime}P^{1}_{n}
-\frac{n}{2{\sqrt{\pi}}}X^{1}_{n}\;,
\;\; (n \ge 1) \nonumber
\eeq
\beq
\alpha^{\perp}_{n}=-{\rm i}{\sqrt{\pi}}\alpha^{\prime}P^{\perp}_{n}
+\frac{n}{2{\sqrt{\pi}}}X^{\perp}_{n}\;,\;\;
\alpha^{\perp}_{-n}={\rm i}{\sqrt{\pi}}\alpha^{\prime}P^{\perp}_{n}
-\frac{n}{2{\sqrt{\pi}}}X^{\perp}_{n}\;,
\;\; (n \ge 1)\;. \nonumber
\eeq
These conditions imply that the canonical 
Hamiltonian \rf{Hamiltonian} vanishes. It 
is possible to write $X^{\mu}(\sigma,\tau)$ directly in
terms of the variables $\alpha^{\mu}_{n}$:
\beq
X^{0}(\sigma,\tau)=X^{0}_{0}(\tau)+{\rm i}{\sqrt{2}}{\sum_{\stackrel{n=-\infty}{n\neq 0}}^{\infty}}\;
\frac{\cos n\sigma}{n}\; \alpha^{0}_{n}(\tau)\;, \nonumber
\eeq
\beq
X^{1}(\sigma,\tau)=\frac{R\sigma}{\pi}+{\rm i}{\sqrt{2}}{\sum_{\stackrel{n=-\infty}{n\neq 0}}^{\infty}}\;
\frac{\sin n\sigma}{n}\; \alpha^{1}_{n}(\tau)\;, \nonumber
\eeq
\beq
X^{\perp}(\sigma,\tau)={\rm i}{\sqrt{2}}{\sum_{\stackrel{n=-\infty}{n\neq 0}}^{\infty}}\;
\frac{\sin n\sigma}{n}\; \alpha^{\perp}_{n}(\tau)\;.  \label{new mode expansion}
\eeq

Invariance under pseudo-conformal transformations $\sigma \rightarrow
\sigma+f(\tau+\sigma)-f(\tau-\sigma)$, $\tau \rightarrow
\tau+f(\tau+\sigma)+f(\tau-\sigma)$, allows for further gauge fixing:
\beq
\partial_{\tau}X^{1}+\partial_{\sigma}X^{0}=0\;,\;\;
\partial_{\tau}X^{0}+\partial_{\sigma}X^{1}=2\alpha^{\prime}E+\frac{R}{\pi}\;,\;\;
X^{0}_{0}=0\;. \label{further}
\eeq
Substituting the expansions \rf{new mode expansion} into \rf{further} gives 
\beq
\partial_{\tau} \alpha^{0}_{n}-{\rm i} n \alpha^{0}_{n}=0 \;, \;\;
\alpha^{0}_{n}+\alpha^{1}_{n}=0\;\;(n\ge 1) \;. \label{algebraic conditions}
\eeq
Substituting \rf{algebraic conditions} into \rf{virasoro}
for $n\neq 0$ yields
\beq
2{\sqrt{2}}\alpha^{\prime}\left(E+\frac{R}{2\pi\alpha^{\prime}} \right)\alpha^{0}_{n}
-\sum_{j+k=n}\alpha^{\perp}_{j}\cdot\alpha^{\perp}_{k}=0\;,\;\;n\neq 0\;, \nonumber
\eeq
or
\beq
\alpha^{0}_{n}=-\alpha^{1}_{n}=\frac{\sqrt{2}}{2\alpha^{\prime}E+\frac{R}{\pi}}
\sum_{j+k=n}\alpha^{\perp}_{j}\cdot\alpha^{\perp}_{k}
\;, \;\;n \neq 0\;.    \label{condition on alpha zero}
\eeq
The conditions \rf{condition on alpha zero} reduce the total number of degrees
of freedom, but do not affect the spectrum of the energy operator and will not be
discussed further. Substituting the algebraic conditions \rf{algebraic conditions} into \rf{virasoro}
for $n=0$ gives
\beq
2\alpha^{\prime}E^{2}-\frac{R^2}{2\pi^{2}}-
\sum_{\stackrel{j=-\infty}{j\neq 0}}^{\infty}\alpha^{\perp}_{j}\cdot\alpha^{\perp}_{-j}
=0\;,   
\eeq
or
\beq
E=\left(\frac{R^{2}}{4\pi^{2}\alpha^{\prime \;2}}+\frac{1}{2\alpha^{\prime \;2}}
\sum_{\stackrel{j=-\infty}{j\neq 0}}^{\infty}\alpha^{\perp}_{j}\cdot\alpha^{\perp}_{-j}
\right)^{\frac{1}{2}}\;. \nonumber
\eeq

Instead of quantizing yet, as was done in reference \cite{arvis}, we 
will first write the
energy $E$ in terms of variables which are functions of $\sigma$. Reintroducing 
the field $X^{\perp}(\sigma, \tau)$ and its conjugate momentum 
${\rm P}^{\perp}(\sigma)$, the energy operator is 
\beq
E=\left\{\frac{R^{2}}{4\pi^{2}\alpha^{\prime \;2}}+\int_{0}^{\pi}\!\!d\sigma
\left[\pi({\rm P}_{\perp})^{2}+
\frac{1}{4\pi \alpha^{\prime\;2}}(\partial_{\sigma}X^{\perp})^{2}\right]
\right\}^{\frac{1}{2}}\;. \nonumber
\eeq
We see that $E^{2}$ contains a piece resembling the Hamiltonian of a free massless
field theory. Upon quantization, we therefore regard
the energy operator as
\beq
E=\left\{\frac{R^{2}}{4\pi^{2}\alpha^{\prime \;2}}+\int_{0}^{\pi}\!\!d\sigma
\left[-\pi
\frac{\delta^{2}}{\delta X^{\perp}(\sigma)\cdot\delta X^{\perp}(\sigma)}+
\frac{1}{4\pi \alpha^{\prime\;2}}(\partial_{\sigma}X^{\perp})^{2}\right]
\right\}^{\frac{1}{2}}\;. \label{new quantized E}
\eeq
Notice that this operator, defined with any sensible regularization is
positive definite.

\setcounter{footnote}{0}

Rewriting \rf{new quantized E} in terms of annihilation operators
$a^{\dagger\;\perp}_{n}=
\frac{{\rm i}}{{\sqrt{n}}\alpha^{\prime}}\alpha^{\perp}_{-n}$
and creation operators $a^{\perp}_{n}=
-\frac{{\rm i}}{{\sqrt{n}}\alpha^{\prime}}\alpha^{\perp}_{n}$, where $n \ge 1$:
\beq
E=\left(\frac{R^{2}}{4\pi^{2}\alpha^{\prime \;2}}+\frac{1}{\alpha^{\prime}}
\sum_{j=1}^{\infty}\;n a^{\dagger \;\perp}_{j}\cdot a^{\perp}_{j}+
\frac{d-2}{2\alpha^{\prime}}\sum_{j=1}^{\infty}\;j
\right)^{\frac{1}{2}}\;. \label{newer quantized E}
\eeq
Unfortunately, this expression presents us with a problem, because $E^{2}$ is
positive definite. We would like 
to make a subtraction $C_{0}$ from $E^{2}$:
\beq
E=\left(\frac{R^{2}}{4\pi^{2}\alpha^{\prime \;2}}+\frac{1}{\alpha^{\prime}}
\sum_{j=1}^{\infty}\;n a^{\dagger \;\perp}_{j}\cdot a^{\perp}_{j}-C_{0}+
\frac{d-2}{2\alpha^{\prime}}\sum_{j=1}^{\infty}\;j
\right)^{\frac{1}{2}}\;. \label{subtracted operator}
\eeq
so that 
\beq
-\frac{2\alpha^{\prime}}{2-d}C_{0}+\sum_{j=1}^{\infty}\;j=\zeta(-1)=-\frac{1}{12}\;,
\label{standard procedure}
\eeq
where $\zeta(s)$ is the Riemann zeta function. Equivalently, the zeta function is
analytically continued to
$s=-1$. This 
would be in accord with Lorentz invariance in the critical dimension 
$D=26$. The same choice follows from requiring the smallest
eigenvalue of $E^{2}$ to be the ``Casimir energy"; the
expression is in quotes, because $E^{2}$ is not actually an 
energy. The smallest eigenvalue of $E$ could then be read off from \rf{subtracted operator}
to give the static potential \cite{alvarez}, \cite{arvis}:
\beq
V(R)=\left(\frac{R^{2}}{4\pi^{2}\alpha^{\prime \;2}}-\frac{d-2}{12\alpha^{\prime}}
\right)^{\frac{1}{2}}=\frac{1}{2\pi \alpha^{\prime}}{\sqrt{R^{2}-R_{c}^{2}}}\;. \label{static potential}
\eeq
For large values of $R$, this has the form \cite{lsw}
\beq
V(R)=\frac{R}{2\pi\alpha^{\prime }}-\frac{\pi(d-2)}{24 R}\;+ \cdots
\;, \nonumber
\eeq
where the leading correction to
the linear potential is universal \cite{universal}. Clearly 
$C_{0}$ is an infinite constant. The 
subtraction
procedure can only be physically meaningful
if (once the world-sheet is suitably regularized) it
can be made by 
introducing a local counterterm in \rf{action}\footnote{We are not disputing
that it is
mathematically sensible to define the zeta function for negative
arguments by analytic continuation. The point is
that there is no physical justification for such a procedure
without introducing an explicit dimensionful cut-off, and a sensible procedure
for making the subtraction. Such a subtraction is impossible for the problem
we are considering, though it is possible in
other situations, e.g. the light-cone gauge string \cite{bn}.}. According 
to our analysis in the next
section, there is {\em no} such procedure. This does not mean that amplitudes
with the constant $C_{0}$ satisfying \rf{standard procedure} introduced are
meaningless; only that these amplitudes are not directly 
obtainable from a quantum action
principle, such as \rf{quantum action principle}.

\section{String evolution and the divergence of string amplitudes} 
\setcounter{equation}{0}
\renewcommand{\theequation}{5.\arabic{equation}}

In this section we begin by calculating the amplitude for a string
to begin in one eigenstate of $X^{\perp}$ at time $X^{0}=0$ and evolve to another
eigenstate of $X^{\perp}$ after a time $X^{0}=T$. This is not quite the same thing as the disk amplitude
with the same boundary conditions;
we propose what the correct form of the latter
should be on the basis of symmetry considerations in Section 7. To illustrate the distinction
between the two quantities, we consider an analogous situation, namely the
difference between the relativistic
particle amplitude and the propagator using similar technology in the appendix.

For any positive operator $D$, the following integral transform will be useful:
\beq
e^{-T\sqrt D}=\frac{T}{\sqrt{\pi}}\int_{0}^{\infty}\frac{du}{u^{2}}e^{-\frac{T^{2}}{4u^{2}}
-u^{2}D}
 \;.\label{integral transform}
\eeq
Hence we expect the amplitude for a string to begin in an eigenstate of the operators
$X^{\perp}(\sigma)$
with eigenvalues $X^{\perp}_{i}(\sigma)$ and finish in an eigenstate with eigenvalues
$X^{\perp}_{f}(\sigma)$ over a time duration $T$ is
\beq
\left< X^{\perp}_{f}(\sigma) \left\vert e^{-TE} \right\vert X^{\perp}_{i}(\sigma) \right>
=\frac{T}{\sqrt{\pi}}\int_{0}^{\infty}\frac{du}{u^{2}}e^{-\frac{T^{2}}{4u^{2}}}
\left< X^{\perp}_{f}(\sigma) \left\vert e^{-u^{2}E^{2}}
\right\vert X^{\perp}_{i}(\sigma) \right> \;, \label{preliminary string}
\eeq
where $E$ is the operator \rf{new quantized E} or \rf{newer quantized E}. If
$E^{2}$ is not positive, the integral fails to converge and
this formula
breaks down. 

To evaluate the matrix element on the right-hand side of 
\rf{preliminary string}, we use the familiar expression for the
simple-harmonic-oscillator kernel:
\beq
\left< q_{f} \left\vert \;e^{-\frac{\beta}{2}\left( 
-\frac{d^{2}}{dq^{2}}+q^{2}
\right)}\; \right\vert q_{i} \right> ={\sqrt{\frac{1}{2\pi \sinh T}}}e^{-A(q_{f},q_{i};\beta)}\;,\nonumber
\eeq
where 
\beq
A(q_{f},q_{i};\beta)=\frac{1}{2} (q_{f}^{2}+q_{i}^{2})\coth \beta -\frac{q_{f}\,q_{i}}{\sinh \beta} 
=\frac{1}{2} \int_{0}^{\beta} dt\; ({\dot q_{\rm class}}^{2}+q^{2}_{\rm class})
\nonumber
\eeq
is the Wick-rotated action 
of the harmonic oscillator for the classical
solution $q_{\rm class}(t)$ for which 
$q_{\rm class}(0)=q_{i}$ and $q_{\rm class}(\beta)=q_{f}$. We thereby obtain
\beq
\left< X^{\perp}_{f}(\sigma) \left\vert e^{-TE} \right\vert X^{\perp}_{i}(\sigma) \right>
=\frac{T}{\sqrt{\pi}}\int_{0}^{\infty}\frac{du}{u^{2}}e^{ -\frac{T^{2}}{4u^{2}} 
-\frac{R^{2}u^{2}}{ 4\pi^{2} \alpha^{\prime \;2} }   }
\;\;\;\;\;\;\;\;\;\;\;\;\;\;\;\;\;\;\;\;\;\;\;\;\;\;\;\;\;\nonumber
\eeq
\beq
\;\;\;\;\;\;\;\;\;\;\;\;\;\;\;\;\;\;\;\;\;\;\;\;\;\;\;\;\;\times
\left[\prod_{n=1}^{\infty} \pi e^{\frac{nu^{2}}{\alpha^{\prime}}}(1-e^{-\frac{2nu^{2}}{\alpha^{\prime}}})
\right]^{-\frac{d-2}{2}}e^{-S_{0}[u; X^{\perp}_{f},X^{\perp}_{i}]}\;,\label{updated string}
\eeq
where
\beq
S_{0}[u; X^{\perp}_{f},X^{\perp}_{i}]=\frac{1}{ 4\pi \alpha^{\prime}}\sum_{n=1}^{\infty}
\int_{0}^{\pi} \!\!d\sigma_{1}\! \int_{0}^{\pi} \!\!d\sigma_{2}\;
n\;\sin n\sigma_{1} \;\sin n\sigma_{2} \;\;\;\;\;\;\;\;\;\;\;\;\;\;\;\;\;\;
\;\;\;\;\;\;\;\;\;\;\; \nonumber
\eeq
\beq
\;\;\;\;\;\;\;\;\;\;\;\;\;\;\;\;\;\;\;\;\;\;\;\;\;\;\;\times
\left\{
[X^{\perp}_{f}(\sigma_{1})^{2}+X^{\perp}_{i}(\sigma_{2})^{2}]
\coth\frac{nu^{2}}{\alpha^{\prime}}
-\frac{2X^{\perp}_{f}(\sigma_{1})\cdot X^{\perp}_{i}(\sigma_{2})}{\sinh\frac{nu^{2}}{\alpha^{\prime}}} \right\}\;. 
\label{classical z action}
\eeq

Recall that the Dedekind eta function of the complex number $\tau$ is 
\beq
\eta(\tau)=e^{-\frac{\pi {\rm i} \tau}{12}}
\prod_{n=1}^{\infty}(1-e^{2\pi {\rm i} n \tau}) \;. \label{inf product}
\eeq
and under an inversion of $\tau$ through the unit circle, followed
by a reflection through the imaginary axis, transforms as follows:
\beq
\eta\!\!\left( -\frac{1}{\tau}\right) ={\sqrt{-{\rm i}\tau}}\;\eta(\tau) \;. 
\label{transformation law}
\eeq
Notice that this function resembles the infinite product in \rf{updated string}. We
write this product as
\beq
\left[\prod_{n=1}^{\infty} \pi e^{\frac{nu^{2}}{\alpha^{\prime}}}(1-e^{-\frac{2nu^{2}}{\alpha^{\prime}}})
\right]^{-\frac{d-2}{2}}\!\!\!
=\eta\!\!\left( \frac{{\rm i} u^{2}}{\pi \alpha^{\prime}}\right)^{-\frac{d-2}{2}}
e^{ -\frac{d-2}{2}\sum_{n=1}^{\infty} \ln \pi } 
e^{-\frac{(d-2)u^{2}}{2\alpha^{\prime}}
(\sum_{n=1}^{\infty}n+\frac{1}{12})
}\; . \nonumber
\eeq
With these manipulations
\beq
\left< X^{\perp}_{f}(\sigma) \left\vert e^{-TE} \right\vert X^{\perp}_{i}(\sigma) \right>
={\cal Z}_{0}\, T \,
\int_{0}^{\infty}\frac{du}{u^{2}}\;\,\eta\!\!\left( 
\frac{{\rm i} u^{2}}{\pi \alpha^{\prime} }
\right)^{-\frac{d-2}{2}} \!\! e^{ -\frac{T^{2}}{4u^{2}} 
-\frac{ R^2 u^{2} }{4\pi^{2} \alpha^{\prime \;2} }     } e^{-S_{0}[u; X^{\perp}_{f},X^{\perp}_{i}]}
\nonumber
\eeq
\beq
\times
\exp \left[ -u^{2}\frac{d-2}{2\alpha^{\prime}}
\left( \sum_{n=1}^{\infty}n+\frac{1}{12}\right)\right]\;,
 \label{nearly final string}
\eeq
where ${\cal Z}_{0}$ is an infinitesimal constant (which can be interpreted as a 
string-wave-function renormalization).

The last factor of
\rf{nearly final string}:  
\beq
\exp -u^{2}\left[\frac{d-2}{2\alpha^{\prime}}
\left( \sum_{n=1}^{\infty}n+\frac{1}{12}\right) \right]\;, \label{last factor}
\eeq
gives a nonsensical result for the amplitude, presenting us with a problem. We 
would like to introduce a subtraction inside the integral on
the right-hand side so that
\rf{last factor} 
can be set equal to unity. Such a procedure mutilates the theory
we began with. If we 
multiply this integral by a factor $e^{C_{0}T}$ we can only
make a subtraction from the energy operator. This is obviously
useless for removing the unwanted factor, even if the limits of integration in $u$
are cut off. We therefore seem to be in a bind; there
is no counterterm which removes this factor. The integral vanishes at large $u$, even
if the world-sheet is regularized. The reason is that any physical
regularization
of the sum $\sum_{n=1}^{\infty}n+\frac{1}{12}$ is greater than 
$R^{2}/4\pi^{2}\alpha^{\prime\;2}$ for sufficiently
large cut-off (though there are formal techniques which obscure the situation). We
believe that this divergence is a fact, and the amplitude 
$$\left< X^{\perp}_{f}(\sigma) 
\left\vert e^{-TE} \right\vert X^{\perp}_{i}(\sigma) \right>$$ is simply
meaningless from a mathematically careful standpoint. 

What we are saying should not be surprising in the light of analytic and numerical
studies of random surface models \cite{adj}. No attempt to define strings with dimension
greater than one has been successful in such studies. As far as we can determine, the 
difficulty is not
related to the tachyon in standard quantization, but simply the impossibility
of making a subtraction.

\section{Formally defined string evolution and the static potential} 
\setcounter{equation}{0}
\renewcommand{\theequation}{6.\arabic{equation}}

At the end
of the last section, we showed that removing the factor \rf{last factor}
leads to an expression which does
not come from the theory of Bosonic 
random surfaces. But let's remove this factor anyway! The answer
is an amplitude consistent with the usual results for the static potential, at
least for $R>R_{c}$. If, as
we expect, but do not prove here, the on-shell extrapolation of such amplitudes
are Veneziano amplitudes, they are still worth investigating. With the unwanted
factor removed from the integrand, the amplitude \rf{nearly final string} becomes
\beq
\raisebox{.6ex}{``} 
\left< X^{\perp}_{f}(\sigma) \left\vert e^{-TE} \right\vert X^{\perp}_{i}(\sigma) \right> \raisebox{.6ex}{"}
={\cal Z}_{0} 
T\,\int_{0}^{\infty}
\frac{du}{u^{2}}
\;\eta\!\!
\left( 
\frac{{\rm i} u^{2}}{\pi \alpha^{\prime}}
\right)^{ -\frac{d-2}{2} }\!\!
e^{ 
-\frac{T^{2}}{4u^{2}} 
-\frac{ R^{2}u^{2} }{  4\pi^{2} \alpha^{\prime\;2}  }
} e^{-S_{0}[u,X^{\perp}_{i},X^{\perp}_{f}]}
\;.\label{final string}
\eeq
The quotation marks mean that this quantity is not really the transition amplitude
for a theory of random surfaces, but instead the expression for the amplitude
with \rf{last factor} removed.

At this point, we shall show that the mode series \rf{classical z action}
for $S_{0}$ may be evaluated. The quantity $S_{0}$ is the Wick-rotated classical action 
\beq
S_{0}[u; X^{\perp}_{f},X^{\perp}_{i}]=\frac{1}{4\pi \alpha^{\prime}}
\int_{0}^{ \frac{u^{2}}{\alpha^{\prime}} }\!\!ds
\int_{0}^{\pi}\!\!d\sigma \;[(\partial_{s}X^{\perp})^{2}+
(\partial_{\sigma}X^{\perp})^{2}]
\;. \label{real space z action}
\eeq
of the 
field $X^{\perp}(\sigma,s)$ with ``time" $s$
the  time and 
initial and final conditions
$$X^{\perp}(\sigma, 0)=X^{\perp}_{i}(\sigma),\;\;
X^{\perp}(\sigma, u^{2}/{\alpha^{\prime}})=X^{\perp}_{f}(\sigma)\;,$$ 
respectively and
the further boundary conditions
$$X^{\perp}(0, s)=X^{\perp}(\pi, s)=0\;.$$ 
In Section 3 we found the solution \rf{classical action} for the classical action with
arbitrary Dirichlet boundary data. It follows from this result that
\beq
S_{0}[u; X^{\perp}_{f},X^{\perp}_{i}]=A[u,X^{\perp}_{i},X^{\perp}_{i}]+
A[u,X^{\perp}_{f},X^{\perp}_{i}]-B[u,X^{\perp}_{i},X^{\perp}_{f}]\;,
\label{profound result1}
\eeq
where $A[u, X^{\perp},Y^{\perp}]$ and
$B[u,X^{\perp},Y^{\perp}]$ are the quadratic forms
\beq
A[u,X^{\perp},Y^{\perp}]=\frac{\pi\alpha^{\prime}}{16(\pi \alpha^{\prime}+u^{2})^{2}}
\int_{0}^{\pi}\!d\sigma_{1}\int_{0}^{\pi}\!d\sigma_{2}\;
\frac{[X^{\perp}(\sigma_{1})-Y^{\perp}(\sigma_{2})]^{2}}
{\sin^{2}\frac{\alpha^{\prime}(\sigma_{1}-\sigma_{2})}{2(\pi\alpha^{\prime}+u^{2}) }}
\;,\label{profound result2}
\eeq
\beq
B[u,X^{\perp},Y^{\perp}]=\frac{\pi\alpha^{\prime}}{8(\pi \alpha^{\prime}+u^{2})^{2}}
\int_{0}^{\pi}\!d\sigma_{1}\int_{0}^{\pi}\!d\sigma_{2}\;
\frac{[X^{\perp}(\sigma_{1})-Y^{\perp}(\sigma_{2})]^{2}}
{\sin^{2}\frac{u^{2}-\alpha^{\prime}(\sigma_{1}+\sigma_{2})}{2(\pi\alpha^{\prime}+u^{2}) }}
\;.\label{profound result3}
\eeq

It is simple to recover 
the static potential $V(R)$ in \rf{static potential} from \rf{final string} for $R>R_{c}$. If
we take $T$ very large, then the integral is dominated by large $u$. For very large
$u$, the quadratic forms \rf{profound result2}, \rf{profound result3} may 
be neglected. The infinite
product formula for the eta function \rf{inf product} 
implies that
\beq
\raisebox{.6ex}{``} 
\left< X^{\perp}_{f}(\sigma) \left\vert e^{-TE} \right\vert X^{\perp}_{i}(\sigma) \right> \raisebox{.6ex}{"}
\longrightarrow {\cal Z}_{0} 
T\,\int_{0}^{\infty}
\frac{du}{u^{2}}
\;e^{
\frac{(d-2)u^{2}}{12\alpha^{\prime}}
}
e^{ 
-\frac{T^{2}}{4u^{2}} 
-\frac{ R^{2}u^{2} }{  4\pi^{2} \alpha^{\prime\;2}  }
} 
={\sqrt{\pi}}{\cal Z}_{0}\,e^{-TV(R)}
\;. \label{large time}
\eeq
This implies that the bulk contribution to the 
free energy of the world sheet is 
$TV(R)$, so that the energy of the string in its 
ground state is $V(R)$. If $R<R_{c}$ we do not find the usual
imaginary result for $V(R)$. In this
case the energy operator has a negative eigenvalue. The
integral over $u$ simply fails to converge. This
means that
\beq
V(R) = \left\{ \begin{array}{c}
\frac{1}{2\pi \alpha^{\prime}} { \sqrt{ R^{2}-R_{c}^{2} }}\;\;\;\;\,\,, \; R>R_{c}   \\
\;\;\;\;\;\;\;\;\;-\infty \;\;\;\;\;\;\;\;\;\;\;\;, \; R<R_{c}
\end{array}  
\right.  \;. \label{final potential}
\eeq
We can regularize the
integral by cutting off the integration variable $u$. As
the cut-off is removed, a negative infinite value
for $V(R)$ is unavoidable. We are not claiming that
references \cite{alvarez}, \cite{arvis}
are incorrect. Our result is a consequence of defining amplitudes with \rf{integral transform}. Under
circumstances when the square of the energy is not positive, the logarithm of
this transform does not become imaginary but infinite.

The amplitude \rf{final string}
is not rotation invariant. For simplicity, assume $X^{\perp}$ on
the boundary is zero. Then $S_{0}=0$. If we take a new integration variable 
$\pi\alpha^{\prime}/u$ and use \rf{transformation law}, \rf{final string} becomes
\beq
\raisebox{.6ex}{``} \left< X^{\perp}_{f}(\sigma) \left\vert e^{-TE} \right\vert X^{\perp}_{i}(\sigma) \right> \raisebox{.6ex}{"}
={\cal Z}_{0} T \int_{0}^{\infty}\frac{du}{u^{2}}
\;(\pi\alpha^{\prime})^{\frac{d-6}{4}}u^{\frac{d+2}{2}}
\;\eta\!\!\left( \frac{{\rm i} u^{2}}{\pi \alpha^{\prime}}\right)^{-\frac{d-2}{2}}\!\!
e^{ -\frac{R^{2}}{4u^{2}} 
-\frac{T^{2}u^{2}}{4\pi^{2}\alpha^{\prime\;2}}}
\;.\nonumber
\eeq
Rotating by $90^{o}$ would yield the same expression, except for the quantity $(\pi\alpha^{\prime})^{\frac{d-6}{4}}u^{\frac{d+2}{2}}$
appearing in the integrand. The lack
of rotation invariance is not really so strange. We are calculating
the amplitude for a string of a specified shape at one time to become
a string with another specified shape at another time. This is
analogous to the amplitude for a 
relativistic
particle to travel from one space-time point
to another space-time point; such an amplitude
is not Lorentz invariant, as shown 
in the appendix. This particle amplitude
and the (Lorentz-invariant) propagator 
have similar integral transforms of type \rf{integral transform}, but with different 
powers of the integration
variable in the integrand. The kernel also has a prefactor
proportional to the time separation, but the propagator
does not.

\section{The off-shell disk amplitude} 
\setcounter{equation}{0}
\renewcommand{\theequation}{7.\arabic{equation}}

In the light of the above considerations, the only sensible form of the 
off-shell disk amplitude in $26$ dimensions contains
an extra factor of $u^{-5}$ and no overall factor of $T$:
\beq
\Phi[T, R ; X^{\perp}_{f}, X^{\perp}_{i} ]
={\cal Z} \,
\int_{0}^{\infty} \frac{du}{u^{7}}
\;\eta\!\!\left( \frac{{\rm i} u^{2}}{\pi \alpha^{\prime}} \right)^{-12}\!\!e^{-\frac{ T^{2} }{ 4u^{2} } 
-\frac{R^{2}u^{2}}{4\pi^{2} \alpha^{\prime\;2}} }e^{-S_{0}[u,X^{\perp}_{i},X^{\perp}_{f}]}
\;, \label{string green's}
\eeq
where ${\cal Z}$ is presumably not the same as ${\cal Z}_{0}$ and $S_{0}$ is given as before by
\rf{profound result1}, \rf{profound result2}, \rf{profound result3}. It is easily checked
that when $X^{\perp}$ vanishes on the boundary the 
right-hand side of \rf{string green's} is 
invariant under rotations by $90^{o}$. Furthermore, the large $T$ behavior
\rf{large time} is not affected. 

The integral expression on the right-hand side of \rf{string green's} suggests 
a guess for the amplitude with
$X^{\perp}$ arbitrary on the boundary (but with $X^{0}$ and $X^{1}$
in the shape of a rectangle of dimensions $T$ and $R$). We use the classical action
$S_{0}$ given in \rf{real space z action}, but without specifying $X^{\perp}=0$
at $\sigma=0,\pi$. The boundary data on the rectangle of dimensions $T\times R$ must be
mapped to boundary data on a rectangle of dimensions $u^{2}/\alpha^{\prime}\times \pi$,
whose perimeter is $2\pi+2\frac{u^{2}}{\alpha^{\prime}}$. We
convert boundary data 
$$X^{\perp}_{i}(\sigma)\;,\;\; X^{\perp}_{\rm left}(X^{0})=X^{\perp}(\sigma=\pi, X^{0})\;,\;\; 
X^{\perp}_{f}(\sigma)\;,\;\; 
X^{\perp}_{\rm right}(X^{0})=X^{\perp}(\sigma=0, X^{0})\;,$$
into 
$X^{\perp}(s)$ with 
$s\in \left[ 0,\;2\pi+2\frac{u^{2}}{\alpha^{\prime}}\right)$ 
by
\beq
X^{\perp}(s)&=&X^{\perp}_{i}(\sigma)\;\;\;\;\;\;,\;\;s=\sigma \;,\;\;\;\;\;\;\;
\;\;\;\;\;\;\;\;\;\;\;\;\;\;\;\;\;\;\;\;\;s\in [0,\;\pi]\;, \nonumber \\
X^{\perp}(s)&=&X^{\perp}_{\rm left}(X^{0})\;\;,\;\; s=\pi+\frac{ u^{2}X^{0} }{\alpha^{\prime} T}\;,\;\;
\;\;\;\;\;\;\;\;\;\;\;\;\;s \in [\pi,\;      \pi+\frac{u^{2}}{          \alpha^{\prime} } ]\;, \nonumber \\
    X^{\perp}(s)
&=& X^{\perp}_{f}(\sigma)\;\;\;\;\;\;,\;\; 
     s=2\pi+\frac{u^{2}}{\alpha^{\prime}}-\sigma\;,\;\;\;\;\;\;\;\;\;\; \;s\in [\pi+\frac{u^{2}}{\alpha^{\prime}},\;2\pi]\;, \nonumber \\
    X^{\perp}(s)
&=& X^{\perp}_{\rm right}(X^{0})\;,\;\;s=2\pi+\frac{2u^{2}}{\alpha^{\prime}}-\frac{u^{2}X^{0}}{\alpha^{\prime} T}
\;,\;\;s\in [2\pi,\; 2\pi+2\frac{u^{2}}{\alpha^{\prime}}] \;. \label{profound result4}
\eeq
This boundary function $X^{\perp}(s)$ can then be used to determine $S_{0}$:
\beq
S_{0}[u,X^{\perp}]=\frac{\pi\alpha^{\prime}}{16(\pi \alpha^{\prime}+u^{2})^{2}}
\oint\!ds_{1}\oint\!ds_{2}\;
\frac{[X^{\perp}(s_{1})-X^{\perp}(s_{2})]^{2}}
{\sin^{2}\frac{\alpha^{\prime}(s_{1}-s_{2})}{2(\pi\alpha^{\prime}+u^{2}) }}
\;,\label{profound result5}
\eeq

Since \rf{string green's} is clearly not the same as \rf{classical action}. The static 
potential obtained by taking $T$ large as in \rf{large time}
is \rf{static potential}. The classical result \rf{classical action}, however, is a simple
area law and therefore leads
to a purely linear potential $V(R)=R/(2\pi\alpha^{\prime})$. The implication of the failure
of the classical result is clear; the quantum fluctuations around the classical solution  
are important. This undoubtably means that the Liouville field does not decouple in twenty-six
dimensions unless Neumann boundary conditions are taken.

\section{Discussion}
\setcounter{equation}{0}
\renewcommand{\theequation}{8.\arabic{equation}}

There are several issues raised in this paper which need to be studied further. The first concerns
the question of what the term  ``off-shell Bosonic
string amplitude" actually means. We have argued that such amplitudes
are not mathematically equivalent 
to those of quantized strings with a Nambu or Polyakov action principle. The question is not simply
a formal one; we would like to know how strings arise effectively in gauge theories. Somehow, these
effective strings are free of the bulk contribution to the square of the energy operator. World-sheet
supersymmetry
may cure this contribution.

Another issue is the determination of off-shell amplitudes for more general boundary
conditions. At present it is not clear how to obtain these directly. It may be possible
to determine string amplitudes of quasi-static strings using our methods. These amplitudes differ
from the static string amplitudes in that the length $R$, of the string is dependent on the 
time $X^{0}$. This may not solve the problem, since the resulting amplitudes will not
be rotation invariant; but perhaps it is possible to guess the answer which generalizes \rf{string green's}.

We remark that it is possible to formally remove the divergence of \rf{string green's} when one
dimension $R$ or $T$ is made sufficiently small. This can be done by simply cutting off the
integration on $u$:
\beq
\Phi[T, R ; X^{\perp}_{f}, X^{\perp}_{i} ]
={\cal Z} \,
\int_{\frac{\pi\alpha^{\prime}}{\Lambda}}^{\Lambda} \frac{du}{u^{7}}
\;\eta\!\!\left( \frac{{\rm i} u^{2}}{\pi \alpha^{\prime}} \right)^{-12}\!\!e^{-\frac{ T^{2} }{ 4u^{2} } 
-\frac{R^{2}u^{2}}{4\pi^{2} \alpha^{\prime\;2}} }e^{-S_{0}[u,X^{\perp}_{i},X^{\perp}_{f}]}
\;, \label{reg string green's}
\eeq
where $S_{0}$ is given by \rf{profound result4}, \rf{profound result5}. This expression is invariant upon interchange of $R$ and $T$. The static potential $V(R)$ is no longer negative-infinite
below the critical distance
$R<R_{c}=\pi{\sqrt{\frac{(d-2)\alpha^{\prime}}{3}}}$, but is instead a finitely-deep potential
well. An interesting question is whether the static potential obtained 
from \rf{reg string green's}
has conceptual or
phenomenological usefulness for QCD.

In both \rf{final string} 
and \rf{string green's} there is a
phase transition at sufficiently small $T$, analogous to the Hagedorn
transition with periodic time. This transition differs from 
the usual Hagedorn
transition in that it is not
associated with windings of the string around a cylinder. The Hagedorn
transition is associated with vortex condensation in the two-dimensional $XY$-model \cite{XY}. The transition of the
world-sheet with rectangular boundary conditions is instead associated with that of the two-dimensional
restricted gaussian model, with partition function
\beq
Z=\left[\prod_{j} \int_{-\beta/2}^{\beta/2} dx_{j}\right] \;\exp-\frac{1}{2}\sum_{<j,k>} (x_{j}-x_{k})^{2}\;, \nonumber
\eeq
where $j$ and $k$ denote sites on a two-dimensional flat lattice and $<j,k>$ means that $j$ and $k$ are
nearest neighbors. This 
lattice model behaves like a massless field theory for sufficiently large $\beta$. It also has
a well-defined high-temperature expansion around $\beta=0$ however, and therefore has a strongly-coupled
phase with exponentially-decaying correlations for small $\beta$. The transition is driven by
the condensation of loops on the lattice where $x=\pm \beta/2$, rather than by vortices.

\section*{Acknowledgement}

I thank Yuri Makeenko for some
insightful remarks on off-shell amplitudes in the critical dimension
which sparked my interest in this topic.

\section*{Appendix: Integral transforms for particle evolution operators and propagators}
\setcounter{equation}{0}
\renewcommand{\theequation}{A.\arabic{equation}}

As an application of the formula \rf{integral transform}, we 
will calculate the wave function of position
${\bf x}$ and time $t$, 
$\Psi({\bf x},t)=K({\bf x}, t;
{\bf x}_{i}, t_{i})$ 
for a relativistic particle which has been restricted to be at initial
position ${\bf x}_{i}$ at initial time
$t_{i}$ (we assume $t>t_{i}$). This  
wave function, the relativistic Schr\"{o}dinger kernel, is 
closely related to the propagator, but they are not
the same, and we discuss the connection between them. Though 
such matters are
elementary, they are neither trivial nor discussed in 
textbooks (at least
to our knowledge), and clarifying
them will facilitate our 
discussion
of the disk amplitude. The particle
amplitude to start from some fixed configuration at one time and finish
in a fixed configuration at a later time is {\em not} the two-point Green's function.

Imagine a relativistic particle scatters isotropically from a small target 
of size $(\Delta x)_{\rm target}$ at ${\bf x}_{i}$
at time $t_{i}$. We measure the probability for the particle to be at $x_{f}$ at time
$t_{f}>t_{i}$ with a small detector of size 
$(\Delta x)_{\rm detect}$ located at $x_{f}$. This probability is
$$(\Delta x)_{\rm target}^{3}(\Delta x)_{\rm detect}^{3}\;\vert K({\bf x}_{f}, t_{f};
{\bf x}_{i}, t_{i}) \vert^{2}.$$ The wave function is the kernel
\beq
K({\bf x}_{f}, t_{f};{\bf x}_{i}, t_{i})= 
\left< {\bf x}_{f} \left\vert\; \exp \left[ \;{\rm i} (t_{f}-t_{i}+
{\rm i} \varepsilon){\sqrt{ -\nabla^{2}+m^{2} }}\; \right] \;
\right\vert {\bf x}_{i}\right> \;, \label{particle wave function}
\eeq
where we have set the speed of light and Planck's constant equal to one
and $m$ is the particle mass.

To evaluate \rf{particle wave function} explicitly, we use \rf{integral transform}
to obtain
\beq
K({\bf x}_{f}, t_{f};{\bf x}_{i}, t_{i})={\rm i} \frac{t_{f}-t_{i}}{\sqrt{\pi}}
\int_{0}^{\infty}\frac{du}{u^{2}}\;e^{\frac{(t_{f}-t_{i})^{2}+{\rm i}\varepsilon}{4u^{2}}}
\left< {\bf x}_{f} \left\vert \;e^{-u^{2}{\sqrt{-\nabla^{2}+m^{2}}}}
\;\right\vert {\bf x}_{i}\right> \;\nonumber
\eeq
\beq
={\rm i}\frac{t_{f}-t_{i}}{8\sqrt{\pi}}
\int_{0}^{\infty}\frac{du}{u^{5}}\;\exp \left[ \frac{(t_{f}-t_{i})^{2}-({\bf x}_{f}-{\bf x}_{i})^{2}
+{\rm i} \varepsilon}{4u^{2}}
-m^{2}u^{2} \right]
\;. \label{integral for w.f.}
\eeq
This integral may be evaluated for 
$\vert (t_{f}-t_{i})^{2}-({\bf x}_{f}-{\bf x}_{i})^{2}\vert
\gg m^{2}$, with the result
\beq
K({\bf x}_{f}, t_{f};{\bf x}_{i}, t_{i}) \simeq i\frac{t_{f}-t_{i}}{8{\sqrt{\pi}}}
\frac{
\int_{0}^{\infty} \frac{du}{u^{5}} \exp  \frac{(t_{f}-t_{i})^{2}
-({\bf x}_{i}-{\bf x}_{f})^{2}+{\rm i} \varepsilon}{4u^{2}}
}{
\int_{0}^{\infty} \frac{du}{u^{2}} \exp  \frac{(t_{f}-t_{i})^{2}
-({\bf x}_{i}-{\bf x}_{f})^{2}+{\rm i} \varepsilon}{4u^{2}}
} \nonumber 
\eeq
\beq
\times \int_{0}^{\infty} \frac{du}{u^{2}}\; \exp \; \left[\frac{(t_{f}-t_{i})^{2}
-({\bf x}_{i}-{\bf x}_{f})^{2}+{\rm i} \varepsilon}{4u^{2}}-m^{2}u^{2}\right]
 \nonumber
\eeq
\beq
=\frac{{\rm i} (t_{f}-t_{i})}{8\pi [(t_{f}-t_{i})^{2}
-({\bf x}_{i}-{\bf x}_{f})^{2}+{\rm i} \varepsilon]^{\frac{3}{2}}}\;
\exp\; {\rm i} m {\sqrt{(t_{f}-t_{i})^{2}
-({\bf x}_{i}-{\bf x}_{f})^{2}+{\rm i} \varepsilon}} \;. \label{result for w.f.}
\eeq

If the proper-time interval is space-like, i.e. $t_{f}-t_{i}<
\vert {\bf x}_{f}-{\bf x}_{i} \vert$, the result \rf{result for w.f.}
shows that the wave function $K({\bf x}_{f}, t_{f};{\bf x}_{i}, t_{i})$ is
not zero, but exponentially decaying. This establishes
that in relativistic quantum mechanics the amplitude for a particle to
travel outside
the light cone does not vanish. This fact is 
almost obvious from the uncertainty principle, but
it is satisfying to understand its origins clearly. Feynman 
showed this result as well, not by an explicit calculation, but 
using a theorem
from harmonic analysis \cite{feyn}. Since the interval between
events is space-like, there exists a Lorentz transformation reversing
the temporal
order of the the scattering and detection of a
particle, which implies the existence of antiparticles (and explains
the PCT theorem). One can even add spin to describe electrons (rather than scalar
particles). To eliminate superluminal 
communication, it is necessary to second quantize (for details see Feynman's article).

As interesting as \rf{result for w.f.} may be, this wave function is not the
same as the propagator. The reason the two are different is that the
the function $K$ is not invariant under a Lorentz transformation. This
can be seen by inspecting our result \rf{result for w.f.}. Yet
a source for a scalar field $j(\bf x)$ is itself a scalar, i.e. is Lorentz
invariant; therefore the propagator must be Lorentz invariant.

The retarded Green's function is
\beq
S_{\rm ret}({\bf x}_{f},t_{f};{\bf x}_{i}, t_{i})=
\left< {\bf x}_{f} \left\vert\;
\frac{\theta(t_{f}-t_{i})}{2\sqrt{-\nabla^{2}+m^{2}}} 
e^{{\rm i} (t_{f}-t_{i}){\sqrt{-\nabla^{2}+m^{2}}}}
\;\right\vert {\bf x}_{i} \right>\;,
\nonumber
\eeq
and the advanced Green's function is
\beq
S_{\rm adv}({\bf x}_{f},t_{f};{\bf x}_{i}, t_{i})=
\left< {\bf x}_{f} \left\vert\;
\frac{\theta(-t_{f}+t_{i})}{2\sqrt{-\nabla^{2}+m^{2}}}
e^{-{\rm i} (t_{f}-t_{i}){\sqrt{-\nabla^{2}+m^{2}}}}
\;\right\vert {\bf x}_{i} \right>\;,
\nonumber
\eeq
where $\theta(t)=0$ for $t<0$ and $\theta(t)=1$ for $t\ge 0$. The 
average of the retarded and advanced 
Green's functions is the scalar-field propagator:
\beq
S({\bf x}_{f},t_{f};{\bf x}_{i}, t_{i})=\frac{1}{2}S_{\rm ret}({\bf x}_{f},t_{f};{\bf x}_{i}, t_{i})
+\frac{1}{2}S_{\rm adv}({\bf x}_{f},t_{f};{\bf x}_{i}, t_{i})\;,
\nonumber
\eeq
or
\beq
S({\bf x}_{f},t_{f};{\bf x}_{i}, t_{i})=
\left< {\bf x}_{f} \left\vert\;
\frac{1}{4\sqrt{-\nabla^{2}+m^{2}}} 
e^{{\rm i}\vert t_{f}-t_{i}\vert{\sqrt{-\nabla^{2}+m^{2}}}}
\;\right\vert {\bf x}_{i} \right>\;. \nonumber
\nonumber
\eeq

The integral transform for the propagator is not \rf{integral for w.f.}, but 
\beq
S({\bf x}_{f},t_{f};{\bf x}_{i}, t_{i})
=-\frac{1}{16\sqrt{\pi}}
\int_{0}^{\infty}\frac{du}{u^{3}}\exp \left[ \frac{(t_{f}-t_{i})^{2}-({\bf x}_{f}-{\bf x}_{i})^{2}
+{\rm i} \varepsilon}{4u^{2}}
-m^{2}u^{2} \right]
\;, \label{integral for prop.}
\eeq
which can be checked by differentiating 
\rf{integral for w.f.} with respect
to $t_{f}$. The right-hand side of equation \rf{integral for prop.} is 
Lorentz invariant. Notice that the power of $u$ in the integral in
\rf{integral for prop.} is different
from that in \rf{integral for w.f.}. It is 
argued in Section 6 that the 
integral expression for the one-closed-string
vacuum expectation value is similar to that for the 
open-string evolution operator, but
has a different power of $u$ in the integrand.

All of the discussion of this appendix has been in 
Minkowski space. We 
could have just as easily done the analysis in 
Euclidean
space, as we do for strings in the main text of this article.

\end{document}